\input harvmac
\noblackbox
%%%%%%%%%%%%%%%%%%%%%%%%%  %%%%%%%%%%%%%%%%%%%%%%%%%%%%%%%
% this set of macros is an extension of harvmac
% (9/91 or later) and its hypertex version, lanlmac.
% for info on hypertex, see
% http://xxx.lanl.gov/hypertex/
%%%%%%%%%%%%%%%%%%% %%%%%%%%%%%%%%%%%%%%%%%%%%%
% determine hypertex mode
%%%%%%%%%%%%%%%%%%%%%%% %%%%%%%%%%%%%%%%%%%%%%%
\newif\ifdraft

\catcode`\@=11
\newif\iffrontpage
%\draft
\newif\ifxxx
\xxxtrue
%\xxxfalse

\newif\ifad
\adtrue
\adfalse

%%%%%%%%%%%%%%%%%%% %%%%%%%%%%%%%%%%%%%%%%%%%%%%%%%%%%%%%%%%%%%%%%
%%%%% figures
%%%%%%%%%%%%%%%%%%% %%%%%%%%%%%%%%%%%%%%%%%%%%%%%%%%%%%%%%%%%%%%%%
\newif\iffigureexists
\newif\ifepsfloaded
\def\epsfcheck{
\ifdraft% to speed up
\input epsf\epsfloadedtrue
\else
  \openin 1 epsf
  \ifeof 1 \epsfloadedfalse \else \epsfloadedtrue \fi
  \closein 1
  \ifepsfloaded
    \input epsf
  \else
\immediate\write20{NO EPSF FILE --- FIGURES WILL BE IGNORED}
  \fi
\fi
\def\epsfcheck{}}
\def\checkex#1{
\ifdraft
\figureexistsfalse\immediate%
\write20{Draftmode: figure #1 not included}
\figureexiststrue %can be commented out. then Figures will not appear
\else\relax
    \ifepsfloaded \openin 1 #1
        \ifeof 1
           \figureexistsfalse
  \immediate\write20{FIGURE FILE #1 NOT FOUND}
        \else \figureexiststrue
        \fi \closein 1
    \else \figureexistsfalse
    \fi
\fi}
\def\missbox#1#2{$\vcenter{\hrule
\hbox{\vrule height#1\kern1.truein
\raise.5truein\hbox{#2} \kern1.truein \vrule} \hrule}$}
\def\lfig#1{%  this is to call the figure in the text
\let\labelflag=#1%
\def\numb@rone{#1}%
\ifx\labelflag\UnDeFiNeD%
{\xdef#1{\the\figno}%
\writedef{#1\leftbracket{\the\figno}}%
\global\advance\figno by1%
}\fi{\hyperref{}{figure}{{\numb@rone}}{Fig.{\numb@rone}}}}
\def\figinsert#1#2#3#4{%  this inserts the figure
\epsfcheck\checkex{#4}%
\def\figsize{#3}%
\let\flag=#1\ifx\flag\UnDeFiNeD
{\xdef#1{\the\figno}%
\writedef{#1\leftbracket{\the\figno}}%
\global\advance\figno by1%
}\fi
\goodbreak\midinsert%
\iffigureexists
\centerline{\epsfysize\figsize\epsfbox{#4}}%
\else%
\vskip.05truein
  \ifepsfloaded
  \ifdraft
  \centerline{\missbox\figsize{Draftmode: #4 not included}}%
  \else
  \centerline{\missbox\figsize{#4 not found}}
  \fi
  \else
  \centerline{\missbox\figsize{epsf.tex not found}}
  \fi
\vskip.05truein
\fi%
{\smallskip%
\leftskip 4pc \rightskip 4pc%
\noindent\ninepoint\sl \baselineskip=11pt%
{\bf{\hyperdef\hypernoname{figure}{{#1}}{}}~}#2%
%{\bf{\hyperdef\hypernoname{figure}{{#1}}{Fig.{#1}}}:~}#2%
\smallskip}\bigskip\endinsert%
}
%%%%%%%%%%%%%%%%%%%%%%%%%%%%%%%%%%%%%%%%%%%%%%%%%%%%%%%%%%%%%%%%%

\def\a{\alpha}
\def\b{\beta}

\def\s{\sigma}
\def\S{\Sigma}

\def\p{\partial}

\def\cd{{\cal D}}
\def\cdb{\bar{\cal D}}
 %gzero
 %gone
 %gtwo
 %gtwo
 %gtwo
 %Rzero
 %Rzero
 %Gammazero
 %Gammazero
 %Gammazero
 %ezero
 %eone
 %etwo
 %ozero
 %oone
 %otwo
 %gzero
 %gone
 %gtwo

\hfill

\def\hb{\hfill\break}

\def\{{\lbrace}
\def\}{\rbrace}

\def\a{\alpha}
\def\b{\beta}
\def\c{\gamma}
\def\da{\dot\alpha}

\def\s{\sigma}

\def\p{\partial}
\def\dim{{\rm dim}}

 %gzero
 %gone
 %gtwo
 %gd
 %hd
 %Rzero
 %nablazero

\def\Box#1{\mathop{\mkern0.5\thinmuskip
           \vbox{\hrule\hbox{\vrule\hskip#1\vrule height#1 width 0pt\vrule}
           \hrule}\mkern0.5\thinmuskip}}
\def\boxx{\displaystyle{\Box{7pt}}}

\def\frac#1#2{{\scriptstyle{#1}\over\scriptstyle{#2}}}

\noblackbox

%%%
\def\abstract#1{
\vskip.5in\vfil\centerline
{\bf Abstract}\penalty1000
{{\smallskip\ifx\answ\bigans\leftskip 2pc \rightskip 2pc
\else\leftskip 5pc \rightskip 5pc\fi
\noindent\abstractfont \baselineskip=12pt
{#1} \smallskip}}
\penalty-1000}
%%%%%%%%%%%%%%%%%%%%%% references %%%%%%%%%%%%%%%%%%%%%%%%%%%%%%
%
\lref\WZ{
J.~Wess and B.~Zumino,
  ``Consequences of Anomalous Ward Identities,''
  Phys.\ Lett.\  B {\bf 37}, 95 (1971).
  %%CITATION = PHLTA,B37,95;%%
}

\lref\BK{
  I.~L.~Buchbinder, S.~M.~Kuzenko,
  ``Ideas and Methods of Supersymmetry and Supergravity: or a Walk Through Superspace,''
    Bristol, UK: IOP (1998) 656 p.}

\lref\AFGJ{
 D.~Anselmi, D.~Z.~Freedman, M.~T.~Grisaru and A.~A.~Johansen,
  ``Nonperturbative Formulas for Central Functions of Supersymmetric Gauge Theories,''
  Nucl.\ Phys.\  {\bf B526}, 543-571 (1998)
  [hep-th/9708042].}

\lref\StT{
S.~Theisen,
  ``Fourth Order Supergravity,''
  Nucl.\ Phys.\  B {\bf 263}, 687 (1986).}
  %%CITATION = NUPHA,B263,687;%%

\lref\Zumino{
B.~Zumino,
``Effective Lagrangians and Broken Symmetries,''
  in `Lectures on Elementary Particles and Quantum Field Theory',
  S. Deser, M. Grisaru, H. Pendleton (eds.),
  M.I.T. Press 1970.}

\lref\tH{
  G.~'t Hooft,
  ``Naturalness, Chiral Symmetry, and Spontaneous Chiral Symmetry Breaking,''
  NATO Adv.\ Study Inst.\ Ser.\ B Phys.\  {\bf 59} (1980) 135.
  %%CITATION = NASBD,59,135;%%
}

\lref\NS{
N.~Seiberg,
  ``Electric - Magnetic Duality in Supersymmetric Non-Abelian Gauge Theories,''
  Nucl.\ Phys.\  B {\bf 435}, 129 (1995)
  [arXiv:hep-th/9411149].
  %%CITATION = NUPHA,B435,129;%%
}

\lref\Cardy{
J.~L.~Cardy,
  ``Is There a c Theorem in Four-Dimensions?,''
  Phys.\ Lett.\  B {\bf 215}, 749 (1988);\hb
  %%CITATION = PHLTA,B215,749;%%
K.~A.~Intriligator and B.~Wecht,
  ``The Exact Superconformal R-Symmetry Maximizes a,''
  Nucl.\ Phys.\  B {\bf 667}, 183 (2003)
  [arXiv:hep-th/0304128].
  %%CITATION = NUPHA,B667,183;%%
}

\lref\PF{
P.~Fayet,
  ``Spontaneous Generation of Massive Multiplets and Central Charges in
  Extended Supersymmetric Theories,''
  Nucl.\ Phys.\  B {\bf 149}, 137 (1979).
  %%CITATION = NUPHA,B149,137;%%
}

\lref\DS{
S.~Deser and A.~Schwimmer,
  ``Geometric Classification of Conformal Anomalies in Arbitrary Dimensions,''
  Phys.\ Lett.\  B {\bf 309}, 279 (1993)
  [arXiv:hep-th/9302047].
  %%CITATION = PHLTA,B309,279;%%
}

\lref\FSBY{
Y.~Frishman, A.~Schwimmer, T.~Banks and S.~Yankielowicz,
  ``The Axial Anomaly and the Bound State Spectrum in Confining Theories,''
  Nucl.\ Phys.\  B {\bf 177}, 157 (1981).
  %%CITATION = NUPHA,B177,157;%%
}

\lref\CG{
S.~R.~Coleman and B.~Grossman,
  ``'t Hooft's Consistency Condition as a Consequence of Analyticity and Unitarity,''
  Nucl.\ Phys.\  B {\bf 203}, 205 (1982).
  %%CITATION = NUPHA,B203,205;%%
}

\lref\RFT{
R.~J.~Riegert,
  ``A Nonlocal Action for the Trace Anomaly,''
  Phys.\ Lett.\  {\bf B134}, 56-60 (1984);\hb
E.~S.~Fradkin and A.~A.~Tseytlin,
  ``Conformal Anomaly in Weyl Theory and Anomaly Free Superconformal Theories,''
  Phys.\ Lett.\  B {\bf 134}, 187 (1984).
  %%CITATION = PHLTA,B134,187;%%
}

\lref\FV{
E.~S.~Fradkin and G.~A.~Vilkovisky,
  ``Conformal Off Mass Shell Extension and Elimination of Conformal Anomalies
  in Quantum Gravity,''
  Phys.\ Lett.\  B {\bf 73}, 209 (1978).
  %%CITATION = PHLTA,B73,209;%%
}

\lref\ST{
A.~Schwimmer and S.~Theisen, in preparation}

\lref\BOU{
N.~Boulanger,
  ``Algebraic Classification of Weyl Anomalies in Arbitrary Dimensions,''
  Phys.\ Rev.\ Lett.\  {\bf 98}, 261302 (2007)
  [arXiv:0706.0340 [hep-th]].
  %%CITATION = PRLTA,98,261302;%%
}

\lref\AC{
  A.~Cappelli, R.~Guida and N.~Magnoli,
  ``Exact Consequences of the Trace Anomaly in Four Dimensions,''
  Nucl.\ Phys.\  B {\bf 618}, 371 (2001)
  [arXiv:hep-th/0103237].
  %%CITATION = NUPHA,B618,371;%%
}

\lref\DDI{
S.~Deser, M.~J.~Duff and C.~J.~Isham,
  ``Nonlocal Conformal Anomalies,''
  Nucl.\ Phys.\  B {\bf 111}, 45 (1976).
  %%CITATION = NUPHA,B111,45;%%
}

\lref\CFL{
 A.~Cappelli, D.~Friedan and J.~I.~Latorre,
  ``C Theorem and Spectral Representation,''
  Nucl.\ Phys.\  B {\bf 352}, 616 (1991).
  %%CITATION = NUPHA,B352,616;%%
}

\lref\DE{
S.~Deser,
  ``Conformal Anomalies: Recent Progress,''
  Helv.\ Phys.\ Acta {\bf 69}, 570 (1996)
  [arXiv:hep-th/9609138]; 
  ``Closed Form Effective Conformal Anomaly Actions in $D\geq 4$,''
  Phys.\ Lett.\  B {\bf 479}, 315 (2000)
  [arXiv:hep-th/9911129].
  %%CITATION = PHLTA,B479,315;%%
  %%CITATION = HPACA,69,570;%%
}

\lref\JOH{
J.~Erdmenger and H.~Osborn,
  ``Conserved Currents and the Energy-Momentum Tensor in Conformally  Invariant
  Theories for General Dimensions,''
  Nucl.\ Phys.\  B {\bf 483}, 431 (1997)
  [arXiv:hep-th/9605009].
  %%CITATION = NUPHA,B483,431;%%
}

\lref\BARVY{
A.~O.~Barvinsky, Yu.~V.~Gusev, G.~A.~Vilkovisky and V.~V.~Zhytnikov,
  ``The One Loop Effective Action and Trace Anomaly in Four Dimensions,''
  Nucl.\ Phys.\  B {\bf 439}, 561 (1995)
  [arXiv:hep-th/9404187].
  %%CITATION = NUPHA,B439,561;%%
}

%%%%%%%%%%%%%%%%%%%%%%%%%%%%%%%%%%%%%%%%%%%%%%%%%%%%%%%%%%%%%%%%%%%

\Title{\vbox{
\rightline{\vbox{\baselineskip12pt}}}}
{Spontaneous Breaking of Conformal Invariance}
\vskip-1cm
{\titlefont\centerline{
{and Trace Anomaly Matching}
\footnote{$^{\scriptscriptstyle*}$}{\sevenrm
Partially supported by GIF,
the German-Israeli Foundation for Scientific Research,
the Minerva Foundation, DIP, the German-Israeli Project Cooperation
and the Einstein Center of Weizmann Institute.}}}
\vskip 0.3cm
\centerline{A.~Schwimmer$^a$ and
S.~Theisen$^b$ }
\vskip 0.6cm
\centerline{$^a$ \it Department of Physics of Complex Systems,
Weizmann Institute, Rehovot 76100, Israel}
\vskip.2cm
\centerline{$^b$ \it Max-Planck-Institut f\"ur Gravitationsphysik,
Albert-Einstein-Institut, 14476 Golm, Germany}
\vskip0.0cm

\abstract{We argue that when conformal symmetry is spontaneously broken
the trace anomalies in the broken and unbroken phases are matched. This puts
strong constraints on the various couplings of the dilaton.
Using the uniqueness of the effective action for the Goldstone supermultiplet
for broken ${\cal N}=1$ superconformal symmetry the dilaton effective action is calculated.}
\Date{\vbox{\hbox{\sl {November 2010}}
}}
\goodbreak

\newsec{Introduction}

The matching of chiral anomalies of the ultraviolet and infrared theories
related by a massive flow plays an important role in
understanding the dynamics of these theories. In particular using the anomaly
matching the  spontaneous breaking of chiral symmetry
in QCD like theories was proven \tH.

For supersymmetric gauge theories  chiral
anomaly matching provides  constraints
when different theories are related by ``non abelian'' duality in the infrared NS.
The matching involves   the equality of a finite number
of parameters, ``the anomaly coefficients'' defined as the values
of certain Green's function at a very special singular point
in phase space. The Green's function themselves have very different structure
at the two ends of the flow.

The massive flows relate by definition conformal theories in the ultraviolet
and infrared  but the trace anomalies of the two theories are not matched: rather the
flow has the property that the $a$-trace anomaly coefficient decreases along it \Cardy.

In this note we study a different set up. We consider a conformal field theory which admits
vacua  where the conformal symmetry is spontaneously broken. A typical example
for such a theory in $d=4$, which we will use as an illustration in the following,
is the Coulomb branch of ${\cal N}=4$ Super-Yang-Mills theory \PF.
Then we can consider the trace anomalies in the broken and unbroken phases respectively.
We will argue that the trace anomalies are matched in the two phases. We stress
that the analytic structure of the amplitudes is different in the two phases:
just the numerical values of the anomalies which depend on the value of the
amplitudes at a very special point are matched.

An essential feature which allows the matching is the validity of the
Ward identities related to the conservation and tracelessness
of the energy-momentum tensor in both phases. This is analogous to the Ward
identities following from global chiral invariance which are valid along the massive flow
and allow the matching of the chiral anomalies.
On the other hand the properties  following from the $SO(d,2)$ conformal
symmetry are not valid in the broken phase since the vacuum is not invariant
under some of the transformations belonging to the conformal group.

In the unbroken phase all the degrees of freedom are massless.
In the broken phase  some of the degrees of freedom become massive.
The trace anomaly in the broken phase will be contributed by the massless degrees of freedom
which include  the ones which were not lifted by the spontaneous breaking
and the ``dilaton'' the Goldstone boson related to the spontaneously broken
conformal symmetry. The massive states play a role in determining through loop effects
the couplings of the dilaton to energy momentum tensors such that the matching occurs.

We will give general arguments for the matching of trace anomalies.
For type A anomalies (we use the terminology of \DS) the argument
is rather similar to the one proving the matching of chiral anomalies \FSBY.
For type B anomalies  a more detailed argument based on an explicit calculation is needed.

When superconformal symmetry is spontaneously broken without breaking the global
supersymmetry a short cut for the proof is available:
since the trace anomalies (both type A and B) are related to chiral
anomalies \AFGJ\ the matching of chiral anomalies implies the matching of
trace anomalies.

The couplings of the dilatons are summarized by the dilaton effective action
in the presence of an external metric which acts as
a source for the energy momentum tensor.
Integrating out the dilaton field we get the generating functional
for the correlators of energy momentum tensors
incorporating the trace anomalies.
It turns out that various generating functionals proposed \RFT\
have analytic properties which identify them as describing
the broken phase. They definitely cannot be used in the unbroken
phase where the analytic structure of the correlators
is different.

In order to get a ``minimal'' effective action for the dilaton
we study spontaneously broken ${\cal N}=1$ superconformal symmetry. In this
case there is a  supermultiplet of Goldstone bosons, the
bosonic components being the dilaton and a Goldstone boson of the spontaneously
broken $U(1)_{\cal R}$ symmetry. The ``minimal''
action for the $U(1)$ boson is well
understood. Therefore, constructing a supersymmetric action for the Goldstone
supermultiplet which reduces to the minimal one
for the  $U(1)$ boson, fixes the dilaton action. This action, after integrating
out the dilaton, gives the generating functional
for energy momentum tensors in the broken phase obeying all the constraints.

The analogue of anomaly matching for a chiral flow where in the
infrared the chiral symmetry is preserved requires that the space
of broken directions (moduli) has more than one point where conformal
symmetry is unbroken.

We do not have a general proof that trace anomalies are matched
when one adds to the space of moduli the space
of truly marginal deformations of the conformal theory though on
specific examples like ${\cal N}=4$ Super Yang-Mills this seems to be
true.

The paper is organized as follows:

In Section 2 we give general arguments for trace anomaly matching and we perform a
calculation for the simplest type B trace anomaly showing the mechanism.

In Section 3 we calculate the ``minimal'' effective action for the Goldstone supermultiplet
for spontaneously broken superconformal supersymmetry.

In Section 4 we discuss the dilaton effective action
following from the supersymmetric action calculated in section 3.
We integrate out the dilaton and we discuss the generating functional obtained,
comparing it to various other actions.
In Section 5 we summarize our results.
We defer the explicit check of trace anomaly matching on the Coulomb branch
of ${\cal N}=4$ supersymmetric Yang-Mills both in the perturbative weak coupling
regime and at strong coupling using the AdS/CFT correspondence to another
publication \ST.

\newsec{Arguments for Trace Anomaly Matching}

A conformal field theory is characterized by an energy momentum tensor $T_{mn}$
which is conserved
\eqn\cons{
\p^m T_{mn}=0}
and traceless
\eqn\tracel{
T^m{}_m=0}
at the quantum level.
Since \cons\ and \tracel\ are operatorial equations of motion, correlators
of energy momentum tensors should obey Ward identities
following from them independently of the vacuum on which the correlators
are evaluated. An unavoidable violation of \tracel\ in certain
correlators gives the trace anomalies.

A very convenient way to study the correlators is to couple
the energy momentum tensor to the perturbation $h_{mn}$ around
flat space of an external metric $g_{mn}$:
\eqn\metr{
g_{mn}=\eta_{mn}+h_{mn}}
Then if we define the generating functional $W(g)$ in this background,
\cons\ is translated into the invariance of $W(g)$ under
diffeomorphisms
\eqn\diffeo{
\delta_\xi g_{mn}=\nabla_m\xi_n+\nabla_n\xi_m}
while \tracel\ into the invariance under Weyl transformations:
\eqn\weyl{
g_{mn}\to e^{2\s(x)}g_{mn}}
If we are interested in correlators involving also other operators,
we couple them to sources, the generating functional depending on the
metric and the sources. We are using the ``passive'' convention i.e. the
space-time coordinates do not transform under symmetry while the metric
and the additional sources have variations at the same space-time point.

The transformations corresponding to the conformal group $SO(d,2)$
are realized as diffeomorphisms on the flat metric compensated by Weyl transformations
such that the flat metric is unchanged. This allows for a clear distinction
between the Ward identities following from \cons,\tracel\ which should be
valid (modulo anomalies) independent of the vacuum i.e. both
in the unbroken and broken phases  and the relations following
from dilations and special conformal transformations, which are obeyed
only in the unbroken phase.

The generating functional should be invariant (modulo
anomalies) under Weyl transformations with parameter $\sigma(x)$
which goes to zero at infinity in both phases since this  implies the validity
of the Ward identities. On the other hand
dilations and special conformal transformations
correspond to diffeomorphisms which do not vanish at infinity and therefore also the Weyl
transformations which compensate them have  a  parameter which does not
vanish at infinity.
Therefore the generating functional should be invariant under Weyl transformations
 which do not vanish at infinity only in the unbroken phase. We will call in the following Weyl
transformations which vanish at infinity ``local''.

In $d=4$ the generating functional $W(g)$ has the possible
trace anomalies:
\eqn\anoo{
\delta_\s W= c\int d^4 x\sqrt{g}\,\s\,C_{mnpq}C^{mnpq}-a \int d^4 x\sqrt{g}\,\s\,E_4}
where $E_4$ is the $d=4$ Euler density and $C_{mnpr}$ is
the Weyl tensor. The anomaly related to the Euler density
is the only type A one while the second one is of type B.
In more than four dimensions the number of type B anomalies
increases and one can have even in $d=4$ type B
anomalies involving other sources.
The dimensionless anomaly coefficients $a$ and $c$ are part
of the characterization of the CFT.

If the conformal symmetry is spontaneously
broken without breaking the Poincar\'e symmetry a Lorentz
scalar primary ${\cal O}$ with dimension $\Delta$ has a vacuum expectation
value:
\eqn\spon{
\langle0|{\cal O}|0\rangle=v^\Delta}
where $v$ has mass dimension $1$. This scale
is the only one in the broken phase characterizing all the dimensionful
quantities appearing.
As a consequence of the Goldstone
theorem there is necessarily a massless Goldstone boson $\tau$,
the dilaton, which couples linearly to the energy momentum
tensor:
\eqn\coupl{ \langle0|T_{mn}|\tau;q\rangle={1 \over 3} f q_m q_n}
where the dilaton $\tau$ is on mass shell with four momentum $q$.
A normalization can be chosen for which $f$ and $v$ are equal and in the following we will
use them interchangeably.

In the broken phase some of the states become massive with masses given
by the scale $v$. This is happening though such that the energy momentum
tensor still obeys \tracel.
The trace anomaly in the broken phase is produced only by the massless states.
These include states which remained massless who contribute through
loops like in the unbroken phase. In addition there is now the dilaton:
its contribution to the anomaly is given by tree diagrams where
the coupling \coupl\ plays an essential role and also
a loop contribution where the dilaton acts like any other
massless scalar.

The statement of anomaly matching is the equality of the $a$ and $c$
coefficients  calculated in the broken phase as described above
with their original values in the unbroken phase.

We now give arguments for the anomaly matching in decreasing
order of generality and (hopefully) increasing rigour.

We start with a general argument: let us calculate the anomaly
coefficients in the broken phase. For a given theory
they can depend only on the scale
parameter $v$ characterizing the breaking. However, $a$ and $c$
being dimensionless, cannot depend on $v$. As a consequence
when $v$ goes to zero and the unbroken phase is recovered, the values of $a$
and $c$ are unchanged. The possible loophole in this argument
is a discontinuity in the limit.

Another general argument valid only for ${\cal N}=1$ superconformal theories uses
the fact \AFGJ\ that the anomalies appear in supermultiplets
and the coefficients $a$ and $c$ are related to chiral
symmetry anomaly coefficients. Since chiral anomalies are matched so will
be also the trace anomaly coefficients.

We give now more specific arguments. We start with the type A anomaly
(the $a$ coefficient). Type A anomalies are very similar
in structure to the chiral anomalies: all the information in $d=4$ is in
the triangle diagram of three energy momentum tensors,
there is no ultraviolet divergence involved and the anomaly
has a topological form which obeys descent equations \BOU. Also
the argument for anomaly matching is very similar to the one for chiral
anomalies \FSBY,\CG.

Consider the correlator of three energy momentum tensors and decompose
it in invariant amplitudes. Since the kinematic decomposition is rather formidable
\AC\ we just give the relevant general structure. The special kinematical
point which is responsible for the anomaly is when all the Lorentz invariants
one can make of the momenta are zero:
\eqn\mom{
q^2=k_1^2=k_2^2=0}
where $q,\,k_1,\,k_2$ are the momenta carried by the energy momentum
tensors. The special point can be approached in different commuting
orders e.g. taking all the invariants to be equal to $q^2$ and then
sending $q^2\to 0$. In this kinematical situation the Ward
identities which follow from \tracel\ and \cons\ reduce to \AC
\eqn\ward{
A(q^2)=0\,,\qquad A(q^2)-q^2 B(q^2)=0}
Here $A$ is an invariant amplitude of dimension zero which multiplies
a kinematical structure corresponding to $E_4$ and $B$ has dimension $-2$.
The amplitude $B$, having negative dimension, is given unambiguously by
\eqn\ano{
B(q^2)={a\over q^2}}
Clearly both Ward identities \ward\ cannot be satisfied simultaneously
and if the second is used to define $A=a$ there will be
a ``type A'' anomaly with coefficient $a$. 

So far the discussion has 
reflected the behaviour around $q^2=0$ and it is equally
valid in the broken and unbroken phases, but the coefficients
are in principle different. From \ano\ we see, however,
that also the high $q^2$ behaviour of the amplitude $B$ is determined.
Since for $q^2\gg v^2$ all effects of the spontaneous breaking
should disappear, the high $q^2$ should match which implies the equality
of the coefficients $a$ in the broken and unbroken phases.

We stress that the correlators in the two phases are very different;
in the unbroken phase all contributions come from loop diagrams (Fig. 1a)
while in the broken phase there is the tree level contribution of the dilaton (Fig. 1b).
In Section 4 we will study in more detail the dilaton couplings
but the contribution of Fig. 1b has the form:
\eqn\tree{
{1\over3}f q_m q_n{1\over q^2}f^\tau E_4}
where a coupling of strength $f^{\tau}$ of the dilaton to two energy momentum tensors
with the kinematical structure of $E_4$ is needed.
Then the contribution of \tree\ to the  coefficient $a$ is ${1\over3} f f^{\tau}$ and knowing $a$ and the other
contributions fixes $f^{\tau}$ in terms of the scale $f$.
\figinsert\figone{\rm Fig. 1:~~ Wavy line represents insertions of energy-momentum
tensor and full line the propagator of a massless degree of freedom
and the dotted the propagator of a dilaton.}{2.0in}{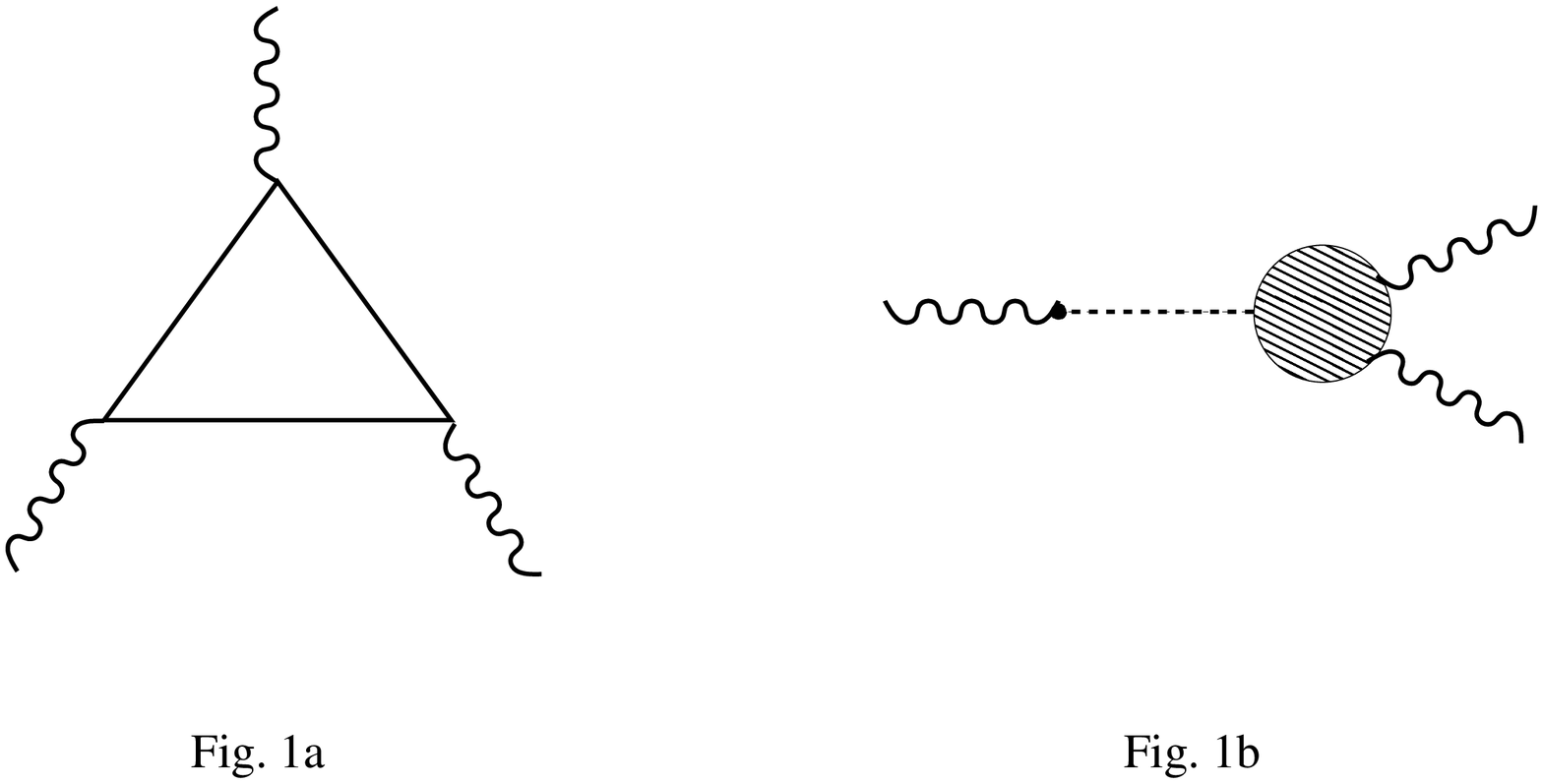}

The structure of \tree, Fig. 1b, reflects the presence of a true pole at $q^2=0$, 
the dilaton, in the broken phase,
while in the unbroken phase the $B$ amplitude \ano\ does not have a pole: the analytic structure of Fig. 1a
is a branch cut which degenerates at the special singular point into an apparent pole, 
but this pole is not present at generic $q^2,\,k_{1}^{2},\,k_{2}^{2}$ and does not correspond
to a state in the Hilbert space.

We now turn to the discussion of the matching for the coefficient $c$.
The previous discussion cannot be generalized in a straightforward fashion
due to the special features of type B anomalies.
In the  unbroken phase  the anomaly  appears in the UV finite part of the three point function
but it is related to the logarithmic divergence of the two
point function \DDI,\CFL. In order to calculate
$c$ one has to put in evidence this connection and therefore one approaches
the singular point through values of the invariants imposed by the two
point function i.e.  $q^2=0$ and $k_{1}^{2}=k_{2}^{2}=k^2$.

On the other hand, in the broken phase the dilaton
contribution does not have a direct connection to
the two point function, the relevant diagram being again Fig. 1b.
Therefore we should put $k^2=0$ at the beginning and keep $q^2\neq 0$ 
in order to put in evidence the dilaton pole.

We will study in detail this mechanism trying again to avoid the very cumbersome
kinematics of the energy momentum three point functions.
We consider the kinematically simplest type B anomaly which singles
out the relevant invariant three and two point amplitudes
which are common to this class of anomalies.

Consider a $d=4$ CFT with a dimension two primary, ${\cal O}^{(2)}$.
This is the case for example in ${\cal N}=4$ Super Yang Mills where the primary
is the bilinear of the scalars in the protected
traceless tensor irrep of $SO(6)$.
The anomaly appears in the
$\langle T_{mn}{\cal O}^{(2)}{\cal O}^{(2)}\rangle$
 correlator.
If we couple ${\cal O}^{(2)}$ to a source $J$ which transforms under
Weyl transformations as:
\eqn\curr{
J(x)\to e^{-2\s(x)}J(x)}
the anomaly has the form:
\eqn\cano{
\delta_\s W(g,J)=\bar c \int d^4 x \sqrt{g}\,\s(x)J^{2}(x)}
Since there is an interplay between the three point function and the correlator of two
${\cal O}^{(2)}$ operators, we expand the generating functional to include these terms:
\eqn\expansion{
W(g,J)=\int d^4 x\,d^4 y\,\Gamma^{(2)}(x,y)\,J(x)\,J(y)
+\int d^4 x\,d^4 y\,d^4 z\,\Gamma^{(3)}_{mn}(x,y,z)\,h^{mn}(x)\,J(y)\,J(z)\,+\,\dots}
and the Ward identities which follow from diffeomorphisms and Weyl
transformations are in momentum space:
\eqn\diff{
q^m\Gamma^{(3)}_{mn}(q,k_1,k_2)={1\over2}(k_1)_n\Gamma^{(2)}(k_1^2)
+{1\over2}(k_2)_n\Gamma^{(2)}(k_2^2)}
\eqn\weyli{
\eta^{mn}\Gamma^{(3)}_{mn}
(q,k_1,k_2)=\Gamma^{(2)}(k_1^2)+\Gamma^{(2)}(k_2^2)}
We have absorbed various factors of $2\pi$ into a convenient normalization.

The kinematical expansion of $\Gamma^{3)}$ is
\eqn\kine{
\Gamma^{(3)}_{mn}=\eta_{mn}\bar A+B q_m q_n+C(q_m r_n+q_n r_m)+D r_m r_n}
with
\eqn\defi{\eqalign{
r_m&=(k_1)_m-(k_2)_m\cr
q_m&=(k_1)_m+(k_2)_m}}
The invariant amplitudes $\bar A,B,C,D$ depend on the invariants $q^2,k_{1}^{2},k_{2}^{2}$.
The two-point function $\Gamma^{(2)}$ depends on just one invariant.

Both $\Gamma^{(2)}$ and $\Gamma^{(3)}$ are UV divergent, however in the combination
\eqn\comb{
A=\bar A-{1\over4}\Big[\Gamma^{(2)}(k_1^2)+\Gamma^{(2)}(k_2^2)\Big]}
the UV divergence is removed. The amplitudes $B,C,D$, having dimension
$-2$, are convergent from the beginning.
We therefore rewrite the Ward identities \diff\ and \weyli\ for this set
of amplitudes:
\eqn\diffone{
A+q^2 B+q\cdot r\, C=0}
\eqn\difftwo{
q^2 C+q\cdot r\, D={1\over4}\Big[ \Gamma^{(2)}(k_1^2)-\Gamma^{(2)}(k_2^2)\Big]}
and
\eqn\weylif{
4 A + q^2 B + 2 q\cdot r\, C + r^2 D=0}
These Ward identities have the generic structure of the relevant identities involving
three and two energy momenta, respectively.

In the unbroken phase there are no singularities at $q^2=0$ for
arbitrary positive $k_{1}^{2},k_{2}^{2}$. Therefore we approach the
singular point choosing also $k_{1}^{2}=k_{2}^{2}=k^2$ such that all
the invariant amplitudes depend only on $k^2$. The Ward
identities \diffone, \difftwo\ and \weylif\ become:
\eqn\diffthree{
A(k^2)=0}
\eqn\difffour{
D(k^2)={1 \over 4} {\p\Gamma^{(2)} \over \p k^2}}
and
\eqn\weylifone{ 4 A(k^2)+4 k^2 D(k^2)=0}
The structure is characteristic for all type B anomalies:
there is a ``clash'' between equations \diffthree, \weylifone\ showing that the anomaly
appears in the three point function, however the normalization
of the anomaly is related to the two point function by equation \difffour.
Explicitly, the two point function in the unbroken phase is
\eqn\two{
\Gamma^{(2)}(k^2)=2\,\bar c\, \log\Big({k^2\over\Lambda^2}\Big)}
where $\Lambda$ is the UV cut off. This leads to:
\eqn\conv{ D(k^2)={\bar c\over 4 k^2}}
the anomaly being $\bar c$, if we choose to put it in the trace, as usual.

We remark for later use that the singular point can also be reached
if we give a mass $m$ to  all the fields $\phi$ in the loop
(the scalar constituents of
the operator ${\cal O}^{(2)}$ in our example) and putting all
invariants to $0$ from the beginning; there are no
infrared singularities for this regulator.
The identity \tracel\ is now modified to:
\eqn\mod{
T^m{}_m=m^2\phi^2}
and, with all kinematical invariants vanishing, \weylif\ takes the form:
\eqn\infr{
4 A(0)=m^2 E(0)}
where $E$, the correlator of two ${\cal O}^{(2)}$ operators and $\phi^2$, is
evaluated at the special point. Dimensionally $E$ must have the form:
\eqn\dim{
E(0)={\bar c\over m^2}}
and the anomaly can be recovered in this way as the limit
of the modified Ward identity when the regulator mass $m$ is sent
to zero. There is a nontrivial relation between the two point function $\Gamma^{(2)}$
and the three point function $E$ which led to
the same $\bar c$, since the limits are interchangeable.

In the above argument and in the following we do not keep track of the various numerical factors
since Fig. 2 shows, as we will explain, how the results match at
the Feynman diagram level.
\figinsert\figone{\rm Fig. 2:~~ The tree diagram for type B. The
full line represents the massive constituent fields
of ${\cal O}^{(2)}$ and the double line
the insertion of ${\cal O}^{(2)}$.}{1.0in}{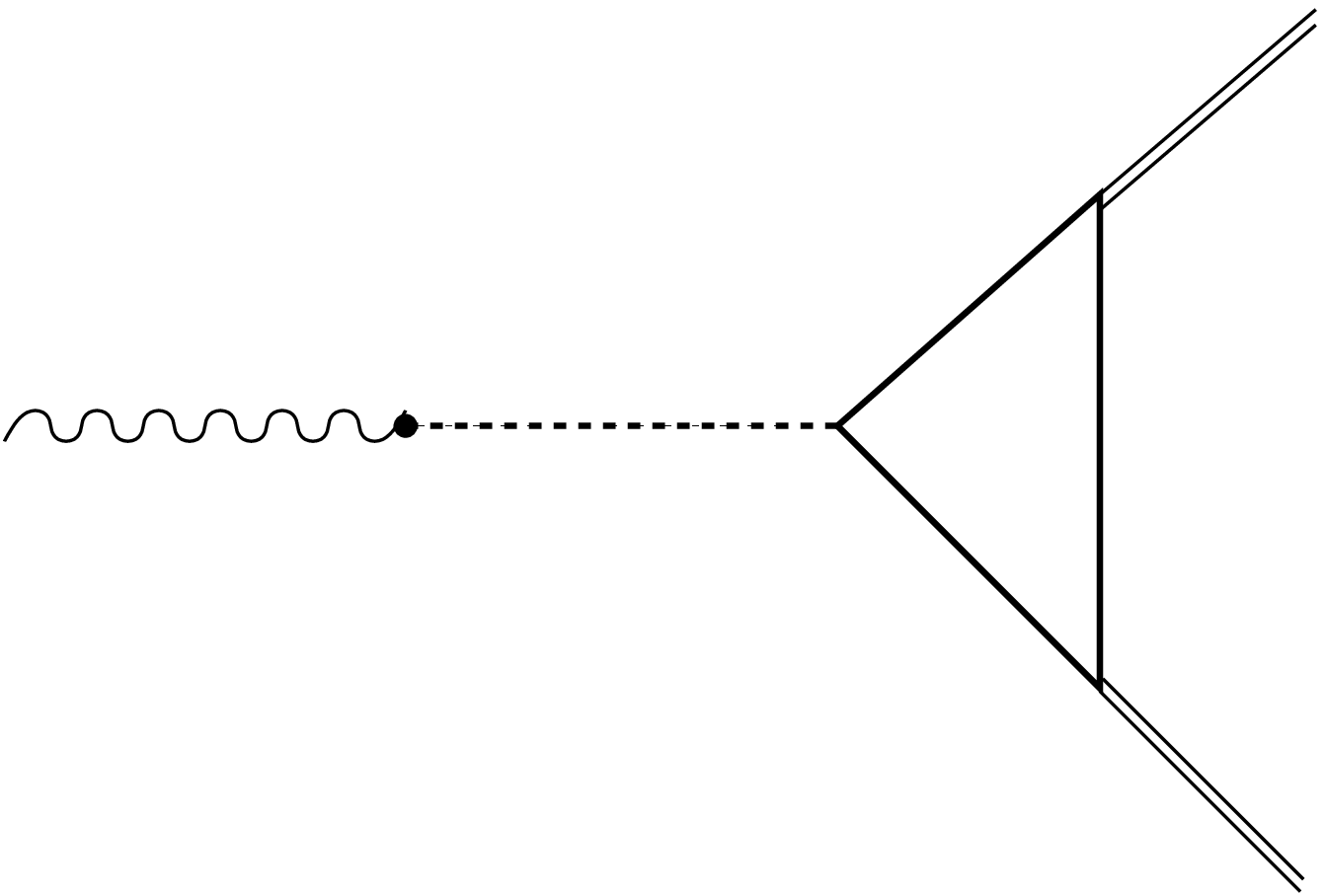}
\noindent

We now evaluate the anomaly in the broken phase.
The Ward identities  \diffone,\difftwo\ and \weylif\ continue to be valid.
If we want to calculate the contribution of the dilaton we need
to evaluate the tree diagram of Fig. 2. Since there is a pole
at $q^2=0$ for any value of the invariants we have to go to the special point
while keeping $q^2$ generic. We do not have a general argument for normalizing
the dilaton coupling to two $J$ sources. We choose therefore a concrete realization
in which the dilaton has a Lagrangian realization, i.e. the  Coulomb branch
of a ${\cal N}=4$ Super Yang-Mills theory with gauge group $SU(2)$.

For concreteness we pick:
\eqn\conc{
{\cal O}^{(2)}\equiv\tr(\phi_5\phi_6)}
and $\phi_1$ as the field which gets expectation value,
the fields $\phi_{i},\, i=1,\dots,6$ being the scalars.

We work in the unitary gauge:
\eqn\gauge{
\phi_1=\pmatrix{\tilde\varphi&\phantom{-}0\cr 0 & -\tilde\varphi}}
Due to our choice of ${\cal O}^{(2)}$ only the part of the Lagrangian
involving the scalars will play a role:
\eqn\lag{
{\cal L}=-\sum_{i=1}^6\tr\Big(D_m\phi_i\,D^m\phi_i\Big)
+g^2\sum_{i<j}\tr\Big(\big[\phi_i,\phi_j\big]^2\Big)}
Since the coupling $g$ is a true parameter in the theory, we can
systematically expand in it. We will calculate the correlator to leading order in
$g$, which is ${\cal O}(g^0)$. The  energy momentum tensor 
of the scalars to order $g^0$ is:
\eqn\enmom{
T_{mn}=-\sum_{i=1}^6\tr\Big[{1\over 6}\p^p\phi_i\p_p\phi_i\,\eta_{mn}-{2\over3}\p_m\phi_i\p_n\phi_i
-{1\over3}\phi_i\boxx\phi_i\,\eta_{mn}+{1\over3}\phi_i\p_m\p_n\phi_i\Big]}
%T_{mn}=\sum_{i=1}^6\tr\Big[\p_m\phi_i\p_n\phi_i-{1\over4}\eta_{mn}\p_p\phi_i\,\p^p\phi_i
%-{1\over2}\phi_i\p_m\p_n\phi_i\Big]}
%
In the broken phase  $\tilde\varphi$ gets an expectation value $v$. The dilaton field is
\eqn\dilaton{
\tau=\tilde\varphi-v}
As a consequence of the breaking, the $\pm$ (in $SU(2)$) components of the scalars 
get massive with a mass
\eqn\mass{
m=gv}
The potential in \lag\ becomes
\eqn\pot{
-g^2\sum_{2\leq i<j\leq6}\tr\Big([\phi_i,\phi_j]^2\Big)
+2 g^2 (\tau+v)^2\sum_{i=2}^6\phi_i^{(+)}\phi_i^{(-)}}
%+16 g^2 v \sum_{i=2}^6\tau\phi_i^{(+)}\phi_i^{(-)}}
%
The contribution of the massive fields to ${\cal O}^{(2)}$ is
\eqn\massi{
\Big(\phi^{(+)}_5\phi^{(-)}_6+\phi^{(-)}_5\phi^{(+)}_6\Big)}
and the energy momentum tensor has the linear term:
\eqn\gold{
-{1\over3}v\,(\p_m\p_n-\eta_{mn}\boxx)\tau}
which corresponds (on-shell) to \coupl.
It is now clear how the matching works since in \pot\
we have the explicit coupling of the dilaton to massive fields
in ${\cal O}^{(2)}$. The massive fields produce an amplitude like $E$
and therefore the coupling of the dilaton to the two currents is
\eqn\coupling{
g^2 v {\bar c\over m^2}=g^2 v {\bar c\over g^2 v^2}={\bar c\over v}}
Even though we used the coupling of the dilaton to the
massive scalars to ${\cal O}(g^2)$, due to the $m^2$ dependence 
the coupling to ${\cal O}^{(2)}$ is ${\cal O}(g^0)$. Together with
the coupling \gold\ the tree diagram Fig. 2 reproduces the
coefficient $\bar c$ of the anomaly. We can see in this example the
general mechanism by which the anomaly is
matched in the broken phase: 
massless field which in the unbroken phase were circulating in the
loop became massive after the breaking. They generate the dilaton
coupling to the operators ${\cal O}^{(2)}$ and since their
contribution
can be viewed alternatively as the calculation of the anomaly
in the unbroken phase with an infrared regulator, the normalizations match.
The numerical factors will obviously match since from Fig. 2 it is evident that the same triangle
gives the normalization of the anomaly in the unbroken phase and fixes the
coupling of the dilaton in the broken phase.

\newsec{The action for the Goldstone supermultiplet
in spontaneously broken  ${\cal N}=1$ superconformal symmetry}

Since there are ambiguities in the dilaton
Lagrangians corresponding to a set of given trace anomalies
we will try to find a ``minimal'' solution which satisfies all the constraints
i.e. reproduces the anomalies and has the right analyticity.

We will do that by starting with a spontaneously broken chiral
$U(1)$ theory. In this case the minimal action for the $U(1)$
Goldstone boson $\beta$ is well understood: coupling the boson to a
$U(1)$ gauge field $A_m$ there is a kinetic term which is gauge
invariant and a term reproducing the chiral anomaly of strength $a$

\eqn\localchiral{
W(A,\beta)=-\int d^4 x\, \Big( (\p_m\b-A_m)(\p^m\b-A^m)
+a\beta F^{mn}\tilde F_{mn}\Big)}

Under gauge transformations, where $A_m\to A_m+\p_m\a$,
the Goldstone field $\beta$ transforms as $\beta\to\beta+\a$
and the variation of \localchiral\ obviously reproduces the
anomaly with coefficient $a$.

Integrating out $\beta$ produces a generating functional depending
on $A_m$ only: the correlators of the $U(1)$ currents having the
correct chiral anomaly follow and the analyticity of this nonlocal
action reflects the broken phase, i.e. a zero mass pole
corresponding to the Goldstone boson.

Now let us consider a spontaneously broken ${\cal N}=1$ superconformal theory
in which also the $U(1)_{\cal R}$
symmetry is spontaneously broken. In such a situation there is a chiral supermultiplet of Goldstone bosons,
the axion $\beta$ and the dilaton $\tau$ appearing together in its lowest component.
If the action for this supermultiplet reduces for $\beta$ to the minimal one, i.e. to 
\localchiral, we will consider the action of $\tau$ also to be ``minimal''.
To achieve this we follow a superspace version of
the Wess-Zumino procedure in order to produce the anomalous term and
then we will add to it a supersymmetry invariant kinetic term.

 We start by reviewing the Wess-Zumino procedure i.e. the construction of the
local action via integration of
the anomaly \WZ. Since it suffices for our purposes, we consider only abelian
symmetries. The Wess-Zumino procedure is trivial in the case of the chiral
$U(1)$, but non-trivial for the Weyl and super-Weyl anomalies.

Consider a quantum field theory which is invariant under an abelian symmetry.
Gauge the symmetry, i.e. couple the QFT to an external gauge field, which is the source
for the symmetry current. Integrating out the quantum fields leads to a non-local
effective action $W(J)$, which is, in general, not invariant under the symmetry,
thus signaling an anomaly ${\cal A}$.
Here $J$ is the external source. Under a symmetry transformation
it transforms to $J^{(\s)}$, where $\s(x)$ is the parameter of the transformation.
If $T$ denotes the generator of the transformation,
\eqn\anomalyone{
(\s T)W(J)=\s\cdot{\cal A}\equiv\int dx\,\s(x)\,{\cal A}(J(x))}
where ${\cal A}$ is a local functional of the source. Let us now assume that
$W(J)$ can be obtained via
\eqn\localaction{
e^{i W(J)}=\int D\phi\, e^{iW(J,\phi)}}
where $W(J,\phi)$ is a local functional of the source and the Goldstone field
$\phi$ which transforms as
\eqn\phitransformation{
\phi\to\phi+\sigma}
such that $J^{(-\phi)}$ is invariant. Applying
\eqn\intid{
e^{\s T}-1=\int_0^1 dt\, \s\,e^{t\s T}}
to $W$ gives
\eqn\Gammaone{
W(J^{(\s)},\phi+\s)=W(J,\phi)+\int_0^1 dt\,\s\cdot{\cal A}(J^{(t\s)})}
The choice $\s=-\phi$ and the normalization $W(J,0)$
lead to
\eqn\GammaJsigma{
W(J,\phi)=\int_0^1 dt\,\phi\cdot{\cal A}(J^{(-t\phi)})}
The existence of $W$ is guaranteed by the following observation:
\eqn\observation{
\phi\,{\cal A}(J^{(-t\phi)})=-{d\over dt}W(J^{(-t\phi)})}
which immediately implies
\eqn\impliciation{
W(J,\phi)=-\Big(W(J^{(-\phi)})-W(J)\Big)}
The r.h.s. should be viewed as a power series in $\phi$. The first
non-vanishing term is the anomaly. All higher order terms are
therefore local.

Note that $W$ is determined up to invariant terms. For
instance, the invariant kinetic energy of the Goldstone field has to
be provided by hand.

We will now turn to the construction of a local effective action,
whose variation under super-Weyl transformation reproduces the
superconformal anomaly. We will do this for ${\cal N}=1$
supersymmetric conformal field theories and as such it also applies
to ${\cal N}=4$ SYM if we impose restrictions, such as $a=c$,  
which are implied by the extended SUSY.

The action $W$ we want to construct depends on the Goldstone  chiral
supermultiplet $\Phi$ coupled to the supergravity gauge multiplet
which acts as a source. The  $U(1)$ gauge and Weyl anomalies belong
now to supermultiplets of anomalies. These supermultiplets contain
also the anomalies of the fermionic supercurrents but since these
are of no interest to us we will put all the fermionic components of
the fields to zero.

The calculation is most easily done in superspace, as we will now explain. Our
discussion uses the notation, conventions and results of \BK\foot{except
for the convention for the Riemann curvature where we use
$[\nabla_m,\nabla_n]V^p=-R_{mn}{}^p{}_q V^q$ which differs by a sign from \BK.}
and for reasons of notational convenience we use in this section a
dimensionless Goldstone superfield. We will restore 
the physical dimensions in the next section.

The variation of the non-local effective action should give the
anomaly
\eqn\vareffaction{\eqalign{
\delta_\Sigma W
&={2(a-c)}\int d^8 z\, {E^{-1}\over R}\delta\Sigma\, W^{\a\b\c}W_{\a\b\c}
\,+\,c.c.\cr
&\qquad-{2c}\int d^8 z\, E^{-1}(\delta\Sigma+\delta\bar\Sigma)(G^a G_a+2 R\bar R)\cr
\noalign{\vskip.2cm}
&=\int d^8z\, {E^{-1}\over R}\delta\Sigma {\delta W\over\delta\Sigma}}}
$a$ and $c$ are the anomaly coefficients.
Some explanation of the notation is in order:
$W_{\a\b\c}$, $R$ (both chiral) and $G_a$ (real) are the gravitational
superfields in terms of which
the solutions of the Bianchi identities can be parametrized. $E$ is the determinant
of the super-vielbein and the integrations are over ${\cal N}=1$ superspace.
$\Sigma$ is a chiral superfield which parametrizes the super-Weyl transformations
under which the various superfields transform as ($G_{\a\da}=\sigma^a_{\a\da}G_a$)
\eqn\Weylb{\eqalign{
R&\to %e^{-2\S}\Big(R-{1\over 4}\cdb^2\Big)e^{\bar\S}=
e^{\bar\S-2\S}\left(R-{1\over4}\Big(\cdb^2\bar\S
+(\cdb_{\da}\bar\S)(\cdb^{\da}\bar\S)\Big)\right)\cr
G_{\a\da}&\to e^{-{1\over2}(\S+\bar\S)}
\left(G_{\a\da}+{1\over2}(\cd_\a\S)(\cdb_{\da}\bar\S)
-i\cd_{\a\da}(\S-\bar\S)\right)\cr
W_{\a\b\c}&\to e^{-{3\over 2}\S}W_{\a\b\c}\cr
E&\to e^{-(\S+\bar\S)}E\cr
\cd_\a&\to e^{{1\over2}\S-\bar\S}(\cd_\a-(\cd^\b \S)M_{\b\a})\cr}}
$M_{\a\b}$ is the Lorentz transformation generator in the $(1/2,0)$ representation.
It acts as
\eqn\actionM{
M_{\a\b}(\psi_\c)={1\over2}(\epsilon_{\c\a}\psi_\b+\epsilon_{\c\b}\psi_\a)}
The infinitesimal version of these are
\eqn\infttrans{\eqalign{
\delta_\S R&=(\bar\S-2\S)R-{1\over4}\cdb^2\bar\S\cr
\delta_\S\bar R&=(\S-2\bar\S)\bar R-{1\over4}\cd^2\S\cr
\delta_\S G_{\a\da}&=-{1\over2}(\S+\bar\S)G_{\a\da}-i\cd_{\a\da}(\S-\bar\S)}}
and
\eqn\inftransDs{\eqalign{
&\delta_\S\cd_\a=\Big({1\over2}\Sigma-\bar\Sigma\Big)\cd_\a\,,\quad
\delta_\S\cdb_{\da}=\Big({1\over2}\bar\Sigma-\Sigma\Big)\cdb_{\da}\cr
&\delta_\S\cd_{\a\da}=-{1\over2}(\S+\bar\S)\cd_{\a\da}
-{i\over2}\Big((\cd_\a\S)\cdb_{\da}+(\cdb_{\da}\bar\S)\cd_\a\Big)\cr
&\delta_\S\cd^2=(\S-2\bar\S)\cd^2+2(\cd^\a\S)\cd_\a\,,\quad
\delta_\S\cdb^2=(\bar\S-2\S)\cdb^2+2(\cdb_{\da}\bar\S)\cdb^{\da}}}
where \inftransDs\ are valid if they act on a Lorentz scalar.
$\cd$ are supercovariant derivatives and 
$\{\cd_\a,\cdb_{\da}\}=-2\,i\,\cd_{\a\da}$.

One can check that \vareffaction\ satisfies the Wess-Zumino consistency
condition
\eqn\WZconsistency{
(\delta_{\S_1}\delta_{\S_2}-\delta_{\S_2}\delta_{\S_1})W=0}
To construct the local effective action, following \WZ, we need the
variation of the anomaly under finite Weyl transformations. Using \Weylb\ we find
\eqn\deltaWZ{\eqalign{
&\Big[(E^{-1}/R)\, W^{\a\b\c}W_{\a\b\c}\Big]^{(\Sigma)}= (E^{-1}/R)\, W^{\a\b\c}W_{\a\b\c}\cr
\noalign{\vskip.2cm}
&\left[E^{-1}\,\Big(G^a G_a+2 R\bar R\Big)\right]^{(\S,\bar\S)}
= E^{-1}\Big\lbrace G_a G^a+2 R\bar R
+i G_{\a\da}\cd^{\a\da}(\S-\bar\S)-{1\over 2}(R\cd^2\S+\bar R\cdb^2\bar\S)\cr
&\qquad-{1\over2}G_{\a\da}(\cd^\a\S)(\cdb^{\da}\bar\S)
-{1\over2}\big(R(\cd\S)^2+\bar R (\cdb\bar\S)^2\big)
+{1\over2}\cd_{\a\da}(\S-\bar\S)\,\cd^{\a\da}(\S-\bar\S)\cr
&\qquad\quad+{1\over8}(\cd^2\S)(\cdb^2\bar\S)
+{i\over2}(\cd_\a\S)(\cdb_{\da}\bar\S)\cd^{\a\da}(\S-\bar\S)
+{1\over8}(\cd^2\S)(\cdb\bar\S)^2+{1\over8}(\cdb^2\bar\S)(\cd\S)^2\Big\rbrace}}
Then the local effective action is
\eqn\localSeff{\eqalign{W_{\rm loc}
&={f^2}\int d^8 z\, E^{-1} e^{-\Phi}e^{-\bar\Phi}\cr
&\quad+{2(c-a)}\int_0^1 dt \int d^8 z\, \Phi
\Big[(E^{-1}/R)W^{\a\b\c}W_{\a\b\c}\Big]^{(-t\Phi)}\,+\,c.c.\cr
&\qquad-{2a}\int_0^1 dt\int d^8 z\,(\Phi+\bar\Phi) \Big[G^a G_a+2
R\bar R\Big]^{(-t\Phi,-t\bar\Phi)}\cr
&={f^2}\int d^8 z\, E^{-1} e^{-\Phi}e^{-\bar\Phi}\cr
&\quad+{2(c-a)}\int d^8 z (E^{-1}/R)\,\Phi\, W^{\a\b\c}W_{\a\b\c}\,+\,c.c.\cr
&\qquad-{2a}\int d^8 z\,E^{-1}\bigg\lbrace(\Phi+\bar\Phi)(G^a G_a+2 R\bar R)
-{1\over2}G_{\a\da}\,\cdb^{\da}\bar\Phi\,\cd^\a\Phi\cr
&\qquad\qquad-{1\over4}\Big(R(\cd\Phi)^2+\bar R(\cdb\bar\Phi)^2\Big)
-{i\over4}\cd^{\a\da}(\Phi-\bar\Phi)\,\cdb_{\da}\bar\Phi\,\cd_\a\Phi
+{1\over16}(\cd\Phi)^2(\cdb\bar\Phi)^2\bigg\rbrace}}
The first line is the super-Weyl invariant kinetic energy term for
the Goldstone field. It is added by hand. To obtain the last two
lines we have repeatedly integrated by parts and used the superspace
Bianchi identities. One may check that the super-Weyl variation of
\localSeff\ is independent of $\Phi$ and reproduces the anomaly.
Here one uses
\eqn\WeylPhi{
\Phi\to\Phi^{(\Sigma)}=\Phi+\Sigma}
We are interested in the effective action in terms of the component
fields. They are defined as\foot{$f$, the auxiliary component of $\S$, should not
be confused with the mass scale $f$ which appears in \coupl.}
\eqn\componentields{\eqalign{
&\Sigma|=S=\s+{2i\over3}\a\,,\qquad -{1\over4}\cd^2\S|=f\cr
&\Phi|=\phi=\tau+i\b\,,\qquad -{1\over 4}\cd^2\Phi|=F\cr
&E^m_a|=e^m_a\,,\qquad R\,|={1\over 3}\bar B\,,\qquad G_a|={4\over3}A_a}}
Here $\Psi|$ means the $\theta=\bar\theta=0$ component of the superfield
$\Psi(z)=\Psi(x,\theta,\bar\theta)$. As before, $\tau$ is the dilaton and $\b$ the axion.
Fermions will be set to zero throughout (which leads to enormous simplifications).

Under (infinitesimal) super-Weyl transformations the component fields transform as
\eqn\transfcomp{\eqalign{
\delta e_m^a=\s e_m^a\,,\quad \delta B=-(\S+3i\a)B-3 f\,,
\quad \delta A_m=\delta(e_m^a A_a)=\p_m\a}}
Furthermore
\eqn\transfPhi{
\delta\phi=S\,,\quad \delta F=f}
We will use the auxiliary component $f$ to transform away $F$. Furthermore,
the auxiliary scalar $B$ is, like $A_m$ and $e_m^a$, an external source which
we can fix by hand. The equation of motion for $B$ gives relation for the auxiliary fields
which never appear linearly in $W_{\rm loc}$. Since we are not interested
in these relations, we will set $B=F=0$ in the component action.

For a chiral superspace Lagrangian ${\cal L}_c$ ($\cdb_{\da}{\cal L}_c=0$),
such as the first term in \vareffaction,
\eqn\compchiral{
\int d^8 z\,{E^{-1}\over R}{\cal L}_c=\int d^4 x \sqrt{g}\Big(-{1\over 4}\cd^2 {\cal L}_c|
+B{\cal L}_c|\Big)}
plus terms which contain the gravitino field. The second term in \compchiral\
obviously vanishes for $B=0$.
For a non-chiral Lagrangian ${\cal L}$, such as the second line in \vareffaction, one uses
that $-{1\over4}(\cdb^2-4R){\cal L}$ is chiral and then \compchiral.

The action in terms of the remaining bosonic component fields is
\eqn\componentaction{\eqalign{
W_{\rm loc}
&={f^2}\int d^4 x\sqrt{g}e^{-2\tau}\Big(-\nabla^m\tau\nabla_m\tau+{1\over6}R
-\big(\nabla_m\b-{2\over3}A_m\big)\big(\nabla^m\b-{2\over3}A^m\big)\Big)\cr
&\quad-\int d^4 x \sqrt{g}\Big[
a\,\tau\,E_4+c\,\tau\,C^{mnpq}C_{mnpq}-{8\over3}\,c\,\tau\, F^{mn}F_{mn}\cr
&\qquad\qquad\qquad\qquad\qquad
+\b\Big({8\over9}\,(3\,c-5\,a)\,F^{mn}\tilde F_{mn}+(a-c)\,R^{mnpq}\tilde R_{mnpq}\Big)\Big]\cr
&\quad+{16\over9}a\int d^4 x\sqrt{g}
\Big(3\Big[-R^{mn}A_n+{1\over6}R A^m+{4\over9} A^2 A^m\Big]\nabla_m\b
+2 A^m A^n\nabla_m\nabla_n\tau\Big)\cr
&\quad-{a}\int d^4 x\sqrt{g}
\Big\{\Big[-\big({R}-{8\over9}A^2\big)g^{mn}+2\big({R}^{mn}-{8\over9}A^m A^n\big)\Big]
\nabla_m\tau\nabla_n\tau\cr
&\qquad+\Big[-\big({1\over3}{R}+{8\over9}A^2\big)g^{mn}
+2\big({R}^{mn}-{8\over9}A^m A^n\big)\Big]\nabla_m\b\,\nabla_n\b
+{16\over3} A^n\nabla^m\tau\,\nabla_n\nabla_m\b\Big\}+\dots}}
Here $R$ is the (torsionless) Ricci scalar (not to be confused with the superfield $R$),
$R_{mn}$ the Ricci tensor, etc., $C_{mnpq}$ the Weyl tensor and
$\sqrt{g}E_4$ is the Euler-density in $d=4$, i.e.
\eqn\Eulerdensity{
E_4=\tilde R^{mnpq}\tilde R_{mnpq}=
R^{mnpq}R_{mnpq}-4 R^{mn}R_{mn}+R^2=C^{mnpq}C_{mnpq}-2 R^{mn}R_{mn}+{2\over3}R^2}
Finally, a tilde denotes dualisation, i.e.
\eqn\dualization{
\tilde R_{mnpq}={1\over2}\,\epsilon_{mn}{}^{st}R_{stpq}\,,\qquad
\tilde F_{mn}={1\over2}\,\epsilon_{mn}{}^{pq}F_{pq}}
The first line are the Weyl and gauge invariant kinetic energies of
the dilaton $\tau$ and of the axion $\b$. The external
metric and the gauge fields play the role of Stueckelberg fields.
The dilaton term can be written in terms of the Weyl-invariant
combination $\tilde g_{mn}=e^{-2\tau}g_{mn}$ as
${1\over6}\int\sqrt{\tilde g}R(\tilde g)$. The second and third
lines contain the anomalies:

The first two terms in the second line
are the type A and type B Weyl anomalies which reflect anomalous
Ward identities in the three point function $\langle TTT\rangle$ of
three energy momentum tensors. The last term in the second line is a
type B Weyl anomaly which is visible in $\langle TJJ\rangle$ where
$J$ is the $U(1)$ vector current. The terms in the third line
reflect anomalous Ward identities in $\langle J_5 JJ\rangle$ and
$\langle J_5 TT\rangle$ where $J_5$ is the axial $U(1)_{\cal R}$
current. They are the chiral and Delbourgo-Salam anomalies,
respectively. Due to supersymmetry all five anomalies are
characterized by just two coefficients, $a$ and $c$.

Note that the fourth line in \componentaction\ is the sum of Weyl and gauge variation
with parameters $\tau$ and $\beta$ of a local term proportional to
\eqn\cohotrivialterm{
\int d^4x\sqrt{g}\Big(R^{mn} A_m A_n-{1\over6}R A^2-{2\over 9}(A^2)^2\Big)}
The second and third lines of \componentaction\ have previously been derived
in \AFGJ. To derive the remaining terms we have also used results in \StT, appropriately
converted to the conventions of \BK.
The terms of third and fourth order in the Goldstone field $S$, which we have not 
written down in \componentaction, 
contain three and four derivatives, respectively.

The $U(1)_{\cal R}$ part of the action reproduces (in a different
normalization of the gauge field) \localchiral. Therefore the
complete part depending only on  the dilaton, which we now display,
is the ``minimal'' dilaton action we were looking for\foot{This is, in fact,
the same action which one obtains from $\int dt\,\tau\cdot\left(c\,[C^2]^{(-t\tau)}-a\,[E_4]^{(-t\tau)}\right)$.}
\eqn\Weffallorder{\eqalign{
W_{\rm loc}&=-{f^2}\int d^4 x\sqrt{g}e^{-2\tau}\Big(\p^m\tau\p_m\tau-{1\over6}R\Big)\cr
&\quad-\int d^4 x\sqrt{g}\Big\{a\tau E_4+c\tau C^{mnpq}C_{mnpq}
+2a\Big(R^{mn}-{1\over2} R g^{mn}\Big)\p_m\tau\p_n\tau\cr
&\qquad\qquad\qquad\qquad+4\,a\,(\p^m\tau\p_m\tau)\,\boxx\tau-2\,a\, (\p^m\tau\,\p_m\tau)^2\Big\}}}
Besides the coupling to the anomalies the action has higher order
terms in the dilaton field. These terms are the analogues of
similar terms in chiral Lagrangians for nonabelian chiral
symmetries, reflecting the nonabelian algebra obeyed by energy
momentum tensors. When  the external metric is flat there is still a
self-interaction of the dilaton: this is the analogue of the
Wess-Zumino-Witten term but the group  being noncompact the term can
be written in a local fashion in $d=4$.

\newsec{The Dilaton Effective Action}

We study now the ``minimal'' dilaton effective action constructed in the previous Section.
After integrating out the dilaton (at  tree level) a generating functional depending only on the
metric is produced. This functional reproduces by construction the anomalies. We will compare
this functional with various other propositions in the literature. The functional corresponds
to the broken phase: it obeys the  Ward identities (modulo anomalies) but it violates  the dilation and/or
special conformal transformation part of the conformal group.

The action for the dilaton has the form:
\eqn\dily{
S=-\int d^4 x\sqrt{g}\Big(e^{-2\tau/f}g^{mn}\p_m\tau\,\p_n\tau-{1\over6}\,f^2 e^{-2\tau/f}R\Big)\,+\,W}
where the first, ``kinetic'' term respects the  Ward identities while the second term
\eqn\anodily{\eqalign{
W=&\,W_a+W_c\cr
=&-a \int d^4 x\sqrt{g}\,\Big\{{1\over f}\,\tau\, E_4\,
+{2\over f^2}\Big(R^{mn}-{1\over2} R\, g^{mn}\Big)\p_m\tau\p_n\tau\cr
&\qquad\qquad\qquad\qquad+{4\over f^3}\,(\p^m\tau\,\p_m\tau)\,\boxx\,\tau
-{2\over f^4}\, (\p^m\tau\,\p_m\tau)^2\Big\}\cr
&-c\int d^4 x\sqrt{g}\,{1\over f}\,\tau\, C^{mnpq}\,C_{mnpq}}}
reproduces the anomalies.
The kinetic term contains the characteristic coupling ${1\over3} f \tau R$
required by Goldstone theorem \coupl.

We start the integration out with the kinetic term. This will also serve to discuss 
the various subtleties which appear in this process.

We first make a change of variable from $\tau$ to a field $\varphi$:
\eqn\change{
\varphi\equiv f\big(1-e^{-\tau/f}\big)}
The kinetic part of the action becomes:
\eqn\kiny{
-\int d^4 x\sqrt{g}\Big(g^{mn}\p_m\varphi\p_n\varphi-{1\over6}R\varphi^2
-{1\over6}f^2 R+{1\over3}f\varphi R\Big)}
We then perform the Gaussian integration over $\varphi$. Alternatively we can think about this step as
calculating the tree diagrams contributed by the vertex linear in $\varphi$. The Jacobian of the transformation
\change\ does not influence the calculation and the result is:
\eqn\zero{
\bar S^{(0)}(g)=\int d^4 x\sqrt{g}R\Phi(g)}
where
\eqn\pphi{
\Phi(g)\equiv{1\over6}f^2\Big[1-{1\over 6\boxx+R}R\Big]}
The functional $\Phi(g)$ was introduced in \FV.
It should be understood, together with similar expressions which we will encounter in the following, 
as a formal expansion in curvatures, representing tree diagrams with dilaton poles. 
$\Phi(g)$ obeys the equation:
\eqn\equa{
(6\boxx+R)\Phi=0}
understood again as holding on the functional expanded in powers of $R$.

We now check the transformation of $\Phi(g)$ and $\bar S^{(0)}$
under Weyl transformations. We have to be careful with boundary
conditions and integration by parts: as we discussed before, the
Ward identities are related to ``local'' Weyl transformations while
the dilation/special conformal transformations require Weyl
transformations which do not vanish at infinity.

We start with the Weyl transformation of $\Phi(g)$. The action in
the various expressions will be from left to right, inverting it
requires an integration by part valid only for transformations which
vanish at infinity. We use the transformations:
\eqn\transf{\eqalign{
\delta_\s R&=-2\s R+6\boxx\s\cr
\delta\boxx&=-2\s\boxx+2\nabla^m\s\,\nabla_m}}
where in the second line it is understood that $\boxx$ acts on a scalar. The
transformation of the propagator and of the propagator acting together
with  $R$ follow:
\eqn\prop{
\delta_\s\Big({1\over\boxx}\Big)=-{1\over\boxx}\delta_\s\boxx{1\over\boxx}={2\over\boxx}\s
-2{1\over\boxx}(\nabla^m\s)\nabla_m{1\over\boxx}}
\eqn\act{
\delta_\s\Big({1\over\boxx}R\Big)=-2{1\over\boxx}(\nabla^m\s)\,\nabla_m{1\over\boxx}R
+6\s-{6\over\boxx}\s\boxx-{12\over\boxx}(\nabla^m\s)\nabla_m}
A general term in the expansion of $\Phi(g)$ transforms as
\eqn\gene{\eqalign{
\big({1\over\boxx}&R\big)^N\delta\big({1\over\boxx}R\big)\big({1\over\boxx}R\big)^M\cr
&=-2\Big\lbrace\big({1\over\boxx}R\big)^N
{1\over\boxx}(\nabla^m\s)\nabla_m\big({1\over\boxx}R\big)^{M+1}
+6\big({1\over\boxx}R\big)^N{1\over\boxx}(\nabla^m\s)\nabla_m\big({1\over\boxx}R\big)^M\Big\rbrace\cr
&\qquad+6\Big\lbrace\big({1\over\boxx}R\big)^N\s\big({1\over\boxx}R\big)^M
-\big({1\over\boxx}R\big)^{N+1}\s\big({1\over\boxx}R\big)^{M-1}\Big\rbrace}}
Writing $\Phi(g)$ as an expansion
\eqn\expa{
\Phi(g)={f^2\over6}\left(1-{1\over6}\sum_{N=0}^\infty\Big(-{1\over6}\Big)^N
\Big({1\over\boxx}R\Big)^{N+1}\right)}
and using \gene, which induces term by term cancellations, we get finally
\eqn\infw{
\delta_\s\Phi=-\s\Phi}
or, for finite transformations,
\eqn\finite{
\Phi\to e^{-\s}\Phi}
The transformation \finite\ cannot be used for a constant $\sigma$
or for any Weyl transformation which does not vanish at infinity.
Using \finite\ we can calculate the variation of $\bar S^{(0)}$:
\eqn\invar{\eqalign{
\delta_\s \bar S^{(0)}&=\int d^4x\sqrt{g}\Big[4\s R\Phi
+(-2\s R+6\boxx\s)\Phi-R\s\Phi\Big]\cr
&=\int d^4 x\sqrt{g}\s(6\boxx+R)\Phi=0}}
where in the last step we used \equa.

Therefore, the first term of the generating functional is invariant
under  ``local'' Weyl transformations: this implies that the Ward
identities are obeyed. On the other hand, 
as it depends on the dimensionful parameter $f$,
it is obviously not invariant under dilations. 

We now include in the integration out the terms which are responsible for the
anomalies, i.e. $W_a$ and $W_c$. After the change of variables \change,
$\tau$ in the $W$s is replaced by  $-f\log(1-\varphi/f)$, understood as a
formal expansion in powers of $\varphi$. We work at tree level, as it
is usually done for the infrared effective actions, treating the
contributions of $W$ as additional vertices:
\eqn\vertic{
\sum_{n=0}^\infty c_n(g)\varphi^n}
We classify the contributions by the number of vertices of type \vertic\ 
which they contain, such that
the term \zero\ corresponds indeed to $\bar S^{(0)}$.

Let us calculate then $\bar S^{(1)}$, i.e. the contribution of
a single ``anomaly vertex''. Each $\varphi$ line in \vertic\ connects
to a $-{1\over3} f\varphi R$ vertex through a $\displaystyle{{1\over 2 [\boxx+R/6]}}$
propagator.
As a result each $\varphi$ in \vertic\ gets replaced by:
\eqn\esss{
{1\over6}f{1\over\boxx+{1\over6}R}R=f\left(1-{6\Phi(g)\over f^2}\right)}
After resumming \vertic\ we get $W$ with the dilaton $\tau$ replaced by:
\eqn\ettt{ -f\log(1-\varphi/f)\to
%-f\log\Big(f+{6\phi\over f^2}-f\Big)=
-f\log\Big({6\Phi\over f^2}\Big)}
Therefore the $\bar S^{(1)}$ contribution is simply:
\eqn\frad{
\bar S^{(1)}(g)=W\Big(\tau=-f\log\Big({6\Phi\over f^2}\Big),g\Big)}
Then $\bar S^{(1)}$ reproduces the anomalies by construction since following from \ettt\ and \finite\
one has the correct transformation rule:
\eqn\trann{
-f\log\Big({6\Phi\over f^2}\Big)\to -f\log\Big({6\Phi e^{-\s}\over f^2}\Big)
=-f\log\Big({6\Phi\over f^2}\Big)+\s f}
needed in the Wess-Zumino procedure.

We calculate now the contribution of diagrams with two vertices of type \vertic,
$c_n(g)\varphi^n$ and $c_m(g)\varphi^m$. One line from each vertex is joined by a propagator while
all the other lines join the $-{1\over3} f \phi R$ vertex. There is a $nm$ symmetry factor.
Resumming all the terms
the symmetry factors can be reabsorbed into a functional derivative with respect
to $\varphi$  while $\varphi$ gets replaced as before by \esss.
We get, therefore, the contribution
\eqn\seccon{
\bar S^{(2)}=\int d^4 x\sqrt{g}\,{1\over\sqrt{g}}{\delta W\over\delta\varphi}
\Big|~{-1\over2{\textstyle\left(\boxx+{1\over6}R\right)}}~
{1\over\sqrt{g}}{\delta W\over\delta\varphi}\Big|}
where
\eqn\repla{
{\delta W\over\delta\varphi}\Big|
\equiv{\delta W\over\delta\varphi}\Big(\varphi=f\Big(1-{6\Phi\over f^2}\Big),g\Big)}
The functional derivative \repla\ can be evaluated by the chain rule:
\eqn\chain{
{\delta W\over\delta\varphi}={\delta W\over\delta\tau}{\delta\tau\over\delta\varphi}
={\delta W\over\delta\tau}{1\over 1-\varphi/f}={f^2\over6}{1\over\Phi(g)}{\delta W\over\delta\tau}}
We now prove that
\eqn\functd{
{\delta W\over\delta\tau}}
is Weyl invariant:
the generator of Weyl transformations
\eqn\weygen{
{\delta\over\delta\s}={\delta\over\delta\tau}+2 g_{mn}{\delta\over\delta g_{mn}}}
commutes with
\eqn\vari{
{\delta\over\delta\tau}}
It follows that
\eqn\invar{
{\delta\over\delta\s}{\delta W\over\delta\tau}={\delta\over\delta\tau}{\delta W\over\delta\s}
={\delta\over\delta\tau}{\cal A}(g)=0}
where we used the fact that the Wess-Zumino construction reproduces
the anomaly, which depends only on the metric. Using \finite\ and  the transformation
\eqn\rtrans{
{1\over6\boxx+R}\to e^{-\s}{1\over 6\boxx+R}e^{3\s}}
it is easy to prove that $\bar S^{(2)}$ is Weyl invariant. A similar treatment shows that all the higher
$\bar S^{(n)}$ are Weyl invariant. We have therefore complete information about the generating functional
associated with the minimal dilaton action $\bar S$. If we decompose it according to the number
$n$ of contributing ``anomaly vertices''
\eqn\genfun{
\bar S(g)=\sum_{n=0}^\infty\bar S^{(n)}(g)}
the $\bar S^{(1)}$ term reproduces the anomalies while all the other
terms are invariant under local Weyl transformations.
The term $\bar S^{(1)}$ was proposed as a generating
functional in  \FV.

All terms have dimensionful parameters and therefore break dilation
invariance. An exception is the three curvature term in $\bar
S^{(1)}$ which, however, does not respect the constraints following
from special conformal transformations. The generating functional
has the correct analyticity for the broken phase, i.e. tree diagrams
with dilaton poles.

The appearance of an infinite number of terms (tree diagrams) is not
surprising: if one integrated out the Goldstone boson field in a
chiral Lagrangian, reproducing the non-abelian chiral anomalies, one
would get rather similar expressions.

We discuss now the relation of this generating functional to the one
introduced in \RFT, which also reproduces
the trace anomalies. We first rederive this Lagrangian starting from an alternative
dilaton effective action.

Consider again the Wess-Zumino procedure, but this time applied to a type A
anomaly which is modified by a cohomologically trivial piece, i.e. replace
$E_4$ by $ E_4-{2\over3}\boxx R$. Then  one obtains an action
(without any additional kinetic part):
\eqn\motto{\eqalign{
\tilde S&=a\int d^4 x\sqrt{g}\Big\{{1\over2}\,\tilde\tau\,\boxx^2\,\tilde\tau
-2\nabla^m\tilde\tau\Big(R_{mn}-{1\over3}g_{mn}R\Big)\nabla^n\tilde\tau
+\tilde\tau\Big(E_4-{2\over3}\boxx R\Big)\Big\}\cr
\noalign{\vskip.2cm}
&\qquad\qquad+c\int d^4 x\sqrt{g}\,\tilde\tau\, C_{mnpq}C^{mnpq}}}
It is easy to check that this action reproduces the anomalies (including the cohomologically
trivial piece of course) with the field $\tilde \tau$ having the transformation:
\eqn\tily{
\tilde\tau\to\tilde\tau+\s}
Interestingly the action has a quadratic form: in this sense one does
not need to add any kinetic term. On the other hand the kinetic term
starts with  a $\boxx^2$ which raises doubts about the analytic
structure of the generating functional. This problem was studied in
\DE. We reanalyze the analytic structure of the RFT  generating
functional \RFT, starting from the dilaton effective action \motto.

We calculate the terms cubic in curvatures which follow from \motto. We treat 
the first term as the kinetic part, which 
gives a $1/(a {\boxx}^{2})$ propagator, and the other terms as interactions.
For the cubic terms one $\boxx$ always cancels and the terms have a ``normal'' analytic structure. For
example, the contribution originating from the terms in \motto\ linear in the curvature
becomes:
\eqn\messy{\eqalign{
&(-2){4\over9}a\,\boxx R\,{1\over\boxx^2}
\mathrel{\mathop \nabla^{\scriptscriptstyle{\leftarrow}}}{}\!^m\,
\Big(R_{mn}-{1\over3}g_{mn}R\Big)\,\nabla^n{1\over\boxx^2}\boxx R\cr
&\qquad =-{8\over9}a\, R{1\over\boxx}
\mathrel{\mathop \nabla^{\scriptscriptstyle{\leftarrow}}}{}\!^m\,
\Big(R_{mn}-{1\over3}g_{mn}R\Big)
\mathrel{\mathop \nabla^{\scriptscriptstyle{\rightarrow}}}{}\!^n\,
{1\over\boxx}R}}
The same cubic terms can be reproduced from
%which do not require a dimensionful constant can be reproduced by
%
%\eqn\corr{\eqalign{
%\int d^4 x\sqrt{g}\Big(\p^m\tau\p_m\tau-{1\over3}\tau R -a\,\tau E_4
%-2\,a\,\big(R^{mn}-{1\over2}R g^{mn}\big)\,\p_m\tau\,\p_n\tau\Big)}}
%
%after obvious rescalings.
%We recognize in \corr\ the terms of the
the ``minimal'' dilaton action \anodily. The ${\cal O}(\tau^3)$
contribute and lead to the relative coefficient $1/3$ between the
two terms in \messy.
%which are responsible for the
%same cubic terms (the term which is third order in $\tau$ changes
%effectively the normalization of the ${\cal O}(\tau^2)$ term).
The cubic
terms which carry the information about the anomalies coincide,
however they are ``completed'' in different ways by the two actions.

If we continue integrating out of the RFT action,
at the four curvature order we find contributions in which the
$\boxx^{-2}$ propagator remains uncancelled. We have to conclude
therefore that the analytic structure of the RFT action is not
compatible with the general requirements, even though the cubic terms
responsible for the anomalies are correctly reproduced.

\newsec{Conclusions}
We presented various arguments that the trace anomaly coefficients are matched in the two phases
of a spontaneously broken conformal field theory.

The matching happens through the contributions of the dilaton in the broken phase replacing
the contributions of the massless states of the unbroken phase which became massive in the
broken one. The matching determines the dilaton couplings to various combinations
of energy momentum tensors in terms of the trace anomaly coefficients characterizing the conformal theory.

The couplings are summarized by the dilaton effective action. A minimal form for this action was
calculated relating it through supersymmetry to the minimal $U(1)$ Goldstone boson action.
Integrating out the dilaton we obtained the generating functional for energy momentum correlators in the broken
phase. This generating functional has the correct and expected analyticity structure, i.e. tree diagrams
with simple Goldstone boson poles.

A natural question to ask is what is the form of the generating functional for the unbroken phase. This action
should have an analytic structure completely different from the ones we discussed, in particular it should not have
any poles. The difference should appear already at the three curvature level since
the terms produced by \messy\ differ from the three energy momentum tensor correlators
calculated in \JOH\ assuming an invariant vacuum i.e. in the unbroken phase. 
The generating functional produced by massless free matter, expanded in powers of the curvature,
was computed in \BARVY.

An explicit check in a nontrivial set up of the trace anomaly matching proposed in this paper can be done
for the Coulomb branch of  ${\cal N}=4$ SUSY Yang-Mills theories. The matching can be checked both at weak
coupling, perturbatively and at strong coupling using AdS/CFT duality\ST.

\bigskip
\noindent
{\bf Acknowledgements}

Very useful discussions with S.~Deser, S.~Elitzur, Z.~Komargodski, 
S.~Kuzenko, D.~Kutasov, E.~Rabinovici, G.~Veneziano and
G.~A.~Vilkovisky are gratefully acknowledged.

\listrefs

\bye